\documentclass[aps,pre,twocolumn,showpacs,floatfix]{revtex4}
\usepackage{amsmath,bm,epsfig}

\usepackage{amsmath,bm,epsfig}

\begin{document}

\title{Multifractality and scale invariance in human heartbeat dynamics}
\author{Emily S. C. Ching} 
\affiliation{Department of Physics and Institute of Theoretical Physics, 
The Chinese University of Hong Kong, Shatin, Hong Kong}
\author{Yue-Kin Tsang}
\affiliation{Scripps Institution of Oceanography, University of California, San
Diego, La Jolla, California 92093-0213, USA}

\date{\today}
 
\begin{abstract}
Human heart rate is known to display complex fluctuations.
Evidence of multifractality in heart rate fluctuations in healthy state has been reported 
[Ivanov et al., Nature {\bf 399}, 461 (1999)]. This multifractal character could be 
manifested as a dependence on scale or beat number of the probability density functions (PDFs) of
the heart rate increments. On the other hand, scale invariance has been recently reported 
in a detrended analysis of healthy heart rate increments 
[Kiyono et al., Phys. Rev. Lett. {\bf 93}, 178103 (2004)]. In this paper,
we resolve this paradox by clarifying that the scale invariance reported
is actually exhibited by the PDFs of the sum of detrended healthy heartbeat intervals taken over
different number of beats, and demonstrating that the PDFs of detrended healthy heart rate
increments are scale dependent. Our work also establishes that this scale invariance is a general
feature of human heartbeat dynamics, which is shared by heart rate fluctuations 
in both healthy and pathological states.
\end{abstract}
\maketitle

\section{Introduction}
The heartbeat interval in human is known to display complex fluctuations,
referred to as heart rate variability (HRV). 
In the past decade, many analyses~\cite{PengPRL1993,Thurner1998,IvanovNature1999,AmaralPRL2001,
Ashkenazy2001,Bernaola2001,Costa2002,Yang2003} have been carried out 
to characterize the statistical features of human HRV, with an aim
to gain understanding of human heartbeat dynamics. An intriguing finding is 
the multifractality in healthy HRV and the loss of 
this multifractal character in pathological HRV in patients with
congestive heart failure~\cite{IvanovNature1999}.
Such multifractal complexity in healthy HRV was further shown to be 
related to the intrinsic properties of the control mechanisms in human heartbeat dynamics 
and is not simply due to changes in external stimulation and 
the degree of physical activity~\cite{AmaralPRL2001}. 

In another complicated phenomenon of fluid turbulence, physical measurements are 
also known to be multifractal~\cite{ParisiFrisch1985}.
In fluid turbulence, it is common to study structure functions,
which are the statistical moments of the increments of the signals at different scales,
and their scaling behavior. Multifractality manifests itself as a 
nonlinear dependence of the scaling exponents on the order of the 
structure functions. This nonlinear dependence is equivalent to the scale dependence 
of the probability density functions (PDFs) of the increments 
of the signals at different scales. These ideas of structure functions in fluid 
turbulence were employed to analyse healthy HRV and 
similar multifractality, with a scale dependence of the PDFs of the heart rate increments 
at different scales or between different number of beats, was indeed found~\cite{LinPRL2001}.
This analogy of human HRV to fluid turbulence was further exploited and
a hierarchical structure found in fluid turbulence~\cite{ShePRL1994} 
was shown to exist also in human HRV, with different parameters for 
heart rate fluctuations in healthy and pathological states~\cite{ChingPRE2004}.
The different values of the parameters can thus be used to 
quantify the multifractal character of healthy HRV and its loss in pathological HRV.
On the other hand, in a recent detrended analysis~\cite{KiyonoPRL2004}
that aims to eliminate the non-stationarity of heartbeat data, 
``scale invariance of the PDFs of detrended healthy heart rate increments" 
was reported, and interpreted as an indication that healthy heartbeat dynamics are in
a critical state.  This finding appears to be in contradiction to the multifractal
character of healthy HRV and needs clarification.

In this paper, we resolve this apparent paradox and further establish that human
heartbeat dynamics exhibit a general scale invariance, which is shared by
heart rate fluctuations in both healthy and pathological states.

This paper is organized as follows.
We first review the statistical character of multifractality in healthy HRV in Sec.~II.
In Sec.~III, we clarify that the scale invariance reported for healthy HRV in 
Ref.~\cite{KiyonoPRL2004} is actually exhibited by the PDFs of the 
sum of detrended heartbeat intervals and demonstrate explicitly that the PDFs of detrended healthy 
heart rate increments are indeed scale dependent and thus consistent with the multifractal character 
of healthy HRV. Then we show that pathological heart rate fluctuations in
patients with congestive heart failure also display this scale invariance.
Our finding thus shows that such scale invariance cannot be an indication of
healthy human heartbeat dynamics being in a critical state. 
In Sec.~IV, we show that this general scale invariance in human heartbeat dynamics
is non-trivial in that it is absent in multifractal
turbulent temperature measurements in thermal convective flows. 
In Sec.~V, we show that the essential effect of the detrended analysis 
is to take out the local mean from the data.
Finally, we summarize and conclude our paper in Sec.~VI.

\section{Statistical signature of multifractality}

For completeness, we first review how the multifractality of healthy human HRV
can be studied using the ideas of structure functions in fluid turbulence.
Consider a dataset of human heartbeat intervals $b(i)$, where $i$ 
is the beat number. The beat-to-beat interval is also known as RR interval
as it is the time interval between successive ``R" peaks in the
electrocardiogram (ECG) time signal. The value of $b(i)$ varies from beat to
beat and this variation is the human HRV. Following the ideas of structure functions
in turbulent fluid flows, one defines~\cite{LinPRL2001} the heart rate
increments between an interval of $n$ beats as,
\begin{equation}
\Delta_n b(i) = b(i+n)-b(i) \ ,
\label{increment}
\end{equation}
which are differences between the heart rate intervals separated by $n$ beats.
The $p$-th order structure functions are the $p$-th order 
statistical moments of the increments:
\begin{equation}
S_p(n) = \langle |\Delta_n b(i)|^p \rangle
\label{structure}
\end{equation}
It was found~\cite{LinPRL2001} that $S_p(n)$ for heart rate fluctuations in healthy state 
exhibits power-law dependence on $n$:
\begin{equation}
S_p(n) \sim n^{\zeta_p}
\label{scaling}
\end{equation}
for $n \approx 8  - 2048$. This power-law or scaling behavior is analogous to 
that found for velocity or temperature structure functions in turbulent fluid flows.
Moreover, the scaling exponents $\zeta_p$ were found to
depend on $p$ in a nonlinear fashion~\cite{LinPRL2001} as in turbulent fluid flows. 
 This implies that for healthy HRV,
the standardized PDFs (with mean substracted then normalized by the standard deviation)
of $\Delta b_n$ changes with the scale or the number of beats $n$, and are thus
scale dependent. In fact when Eq.~(\ref{scaling}) holds, 
the standardized PDFs of $\Delta b_n$ are scale invariant, {\it i.e.}, 
independent of $n$, if and only if $\zeta_p$ is
proportional to $p$. Hence the nonlinear dependence of $\zeta_p$ on $p$ or,
equivalently, the scale dependence of the standardized PDFs of $\Delta b_n$ on $n$
is a characteristic signature of the multifractality of healthy HRV, in analogy to
turbulent fluid flows. 

\section{Scale invariance in the detrended analysis}

In contrast to physical measurements in turbulent fluid flows, human heartbeat
interval data are often non-stationary. This non-stationarity is one possible reason for the 
relatively poor quality of scaling in HRV as compared to that in 
turbulent fluid flows.  To eliminate the
non-stationarity, a ``detrended fluctuation analysis" has been
introduced~\cite{PengChaos1995}. This analysis was further developed to study
detrended heart rate increments~\cite{KiyonoPRL2004,KiyonoIEEE2006}. 
The procedure of this detrended analysis consists of the following steps.
First, $B(m)$, which is the sum of $b(j)$:
\begin{equation}
B(m) = \sum_{j=1}^m b(j) \ , 
\label{Bm}
\end{equation}
is calculated.
Second, the dataset of $B(m)$ is divided into segments of size $2n$, 
and the datapoints in each segment is fitted by the best $q$th-order polynomial. 
This polynomial fit represents the ``trend" in the corresponding segment.
Third, these polynomial fits are subtracted from $B(m)$ to get $B^*(m)$, which are
then ``detrended". Finally, the standardized PDFs of the increments of $B^*$:
\begin{equation}
\Delta_n B^*(i) = B^*(i+n)-B^*(i)
\label{DeltaB*}
\end{equation}
for different values of 
$n$ are studied. Note that in Refs.~\cite{KiyonoPRL2004,KiyonoIEEE2006},
$\Delta_n B^*(i)$ was denoted as $\Delta_n B(i)$ 
and the symbol $s$ was used in place of $n$.
The standardized PDFs of $\Delta_n B^*(i)$ for healthy heartbeat data
were found to be independent of $n$, and this was referred to
as ``scale-invariance in the PDFs of detrended healthy human heart rate increments" 
in Ref.~\cite{KiyonoPRL2004}. We shall show below that this conclusion is inaccurate.

Let us denote the detrended heartbeat interval by $b^*(i)$. From the detrended procedure
described above, it is natural to define $b^*(i)$ by
\begin{equation}
B^*(m) = \sum_{j=1}^m b^*(j)
\label{defb*}
\end{equation}
Then 
\begin{equation}
\Delta_n B^*(i) = \sum_{j=i+1}^{i+n} b^*(j)
\label{deltaB*}
\end{equation}
Thus the detrended heart rate increment between $n$ beats
should be defined as
\begin{equation}
\Delta_n b^*(i) = b^*(i+n)-b^*(i)
\label{delatb*}
\end{equation}
Hence $\Delta_n B^*(i)$ is the {\it sum} of detrended heartbeat intervals taken 
over $n$ beats rather than detrended heart rate increments.
As a result, the observation of scale-invariant or
$n$-independent standardized PDFs of $\Delta_n B^*$ does not necessarily
imply that the standardized PDFs of $\Delta_n b^*$ are also $n$-independent.
Indeed, one expects the contrary, namely, the standardized PDFs of 
$\Delta_n b^*$ should depend on $n$ as healthy human HRV is multifractal.

\begin{figure}[hbt]
\centering
\includegraphics[width=.45\textwidth]{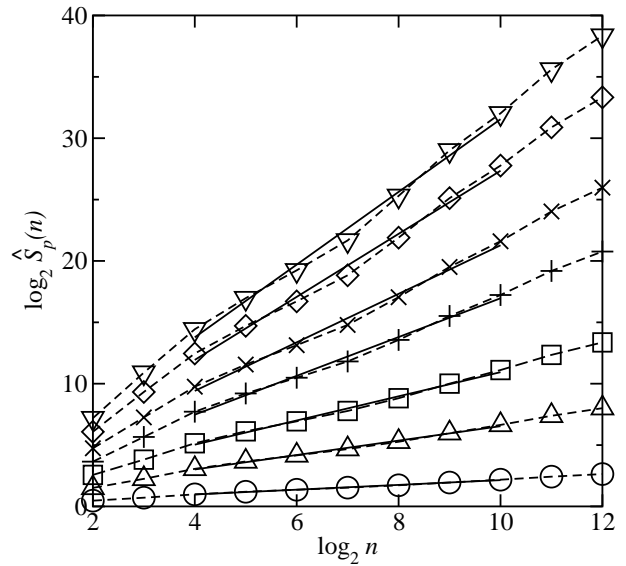}
\caption{The statistical moments $\hat{S}_p(n)$ of the sum of detrended heartbeat
intervals [see Eq.~(\ref{structureB*}) for definition] for healthy heartbeat data 
for $p=0.2$~(circles), $p=0.6$~(triangles), $p=1.0$~(squares), $p=1.6$~(plusses),
$p=2.0$~(crosses), $p=2.6$~(diamonds), and $p=3.0$~(inverted triangles). 
The curves have been shifted vertically for clarity.}
\label{fig1}
\end{figure}

\begin{figure}[thb]
\centering
\includegraphics[width=.45\textwidth]{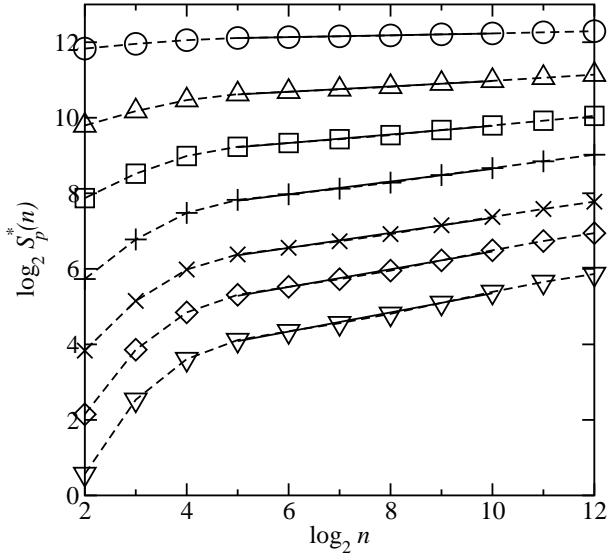}
\caption{The statistical moments $S_p^*(n)$ of detrended heart rate intervals
[see Eq.~(\ref{structureb*}) for definition] for healthy heartbeat data 
for $p$ ranges from 0.2 to 3.0. Same symbols as in Fig.~1. 
The curves have been shifted vertically for clarity.}
\label{fig2}
\end{figure}

To investigate this issue, we study the scaling behavior of the statistical moments
of $\Delta_n B^*$ and $\Delta_n b^*$. As the standardized PDFs of 
$\Delta_n B^*$ are $n$ independent, the scaling exponents for the statistical moments
of $\Delta_n B^*$ should be proportonal to the order of the moments.
On the other hand, we expect the scaling exponents for the statistical moments of 
$\Delta_n b^*$ to have a nonlinear dependence on the order of the moments.
We analyze healthy heartbeat data that are taken from a database of 18 sets of daytime 
normal sinus rhythm data downloaded from public domain\cite{physionet}. 
We follow the detrended procedure described above to get $\Delta_n B^*(i)$. 
We find that a polynomial of degree 3 ($q=3$) is sufficient to
fit the ``trend" as found in Ref.~\cite{KiyonoPRL2004}.
To get the detrended heart rate increment $b^*(i)$,
we use Eq.~(\ref{defb*}) to get
\begin{equation}
b^*(i) = B^*(i)-B^*(i-1)
\label{b*}
\end{equation}
for both $B^*(i-1)$ and $B^*(i)$ belonging to the same segment and skip
that datapoint when $B^*(i-1)$ and $B^*(i)$ fall into different (consecutive)
segments. Next, we evaluate the statistical moments 
\begin{eqnarray}
\hat{S}_p(n) &\equiv& \langle |\Delta_n B^*(i)|^p \rangle
\label{structureB*}\\
S_p^*(n) &\equiv& \langle |\Delta_n b^*(i)|^p \rangle 
\label{structureb*}
\end{eqnarray}
As seen from Fig.~\ref{fig1}, $\hat{S}_p(n)$ exhibits 
power-law or scaling behavior with $n$ with exponents $\hat{\zeta}_p$:
\begin{equation}
\hat{S}_p(n) \sim n^{\hat{\zeta}_p}
\label{hatzeta}
\end{equation}
for $n$ between 16 to 1024 and $p$ between 0.2 to 3. 
On the other hand, $S_p^*(n)$ exhibits better scaling behavior with $n$
with exponents $\zeta^*_p$:
\begin{equation}
S_p^*(n) \sim n^{\zeta^*_p}
\label{zeta*}
\end{equation}
for $n$ between 32 to 1024 and $p$ between $0.2$ to $3$ (see Fig.~\ref{fig2}).

\begin{figure}[thb]
\centering
\includegraphics[width=.45\textwidth]{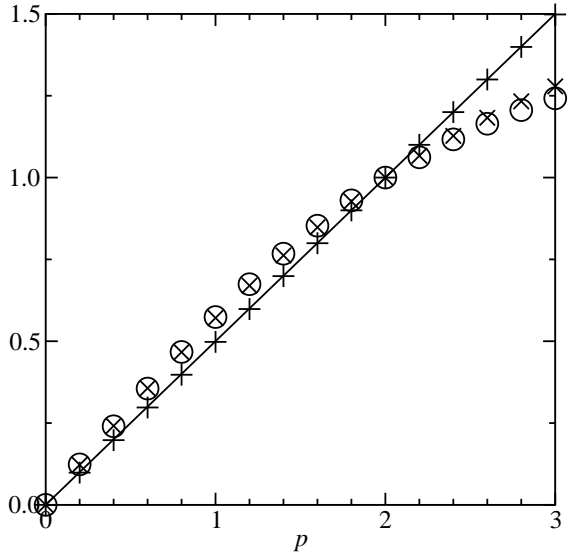}
\caption{The exponents $\hat{\zeta}_p/\hat{\zeta}_2$~(plusses), 
$\zeta^*_p/\zeta^*_2$~(crosses), and $\zeta_p/\zeta_2$~(circles)
as a function of $p$ for healthy heartbeat data. It can be seen that
$\hat{\zeta_p}/\hat{\zeta_2}$ is close to $p/2$ which is shown as the solid line.}
\label{fig3}
\end{figure}

In Fig.~\ref{fig3}, we plot the scaling exponents $\hat{\zeta}_p$ and
$\zeta^*_p$ as a function of $p$. It can be seen that $\hat{\zeta}_p$ is
proportional to $p$ confirming that the standarized PDFs of $\Delta_n B^*$
are scale invariant as reported in~Ref.\cite{KiyonoPRL2004}. On the other hand,
$\zeta^*_p$ is not proportional to $p$ but changes with $p$ in a nonlinear manner,
as expected from the multifractal character of healthy human HRV.
Thus we have clarified that for healthy human HRV that is multifractal,
the standardized PDFs of detrended heart rate increments between $n$ beats depend on $n$
while those of the sum of detrended heartbeat
intervals taken over $n$ beats are $n$-independent.
 We have also calculated the scaling exponents $\zeta_p$ of $S_p(n)$ of the statistical
moments of untreated heart rate increments and the results are shown 
in Fig.~\ref{fig3} too. It can be seen that the detrended procedure does not
change much the scaling exponents of the heart rate increments. We shall return to
this in Sec.~V.

\begin{figure}[thb]
\centering
\includegraphics[width=.45\textwidth]{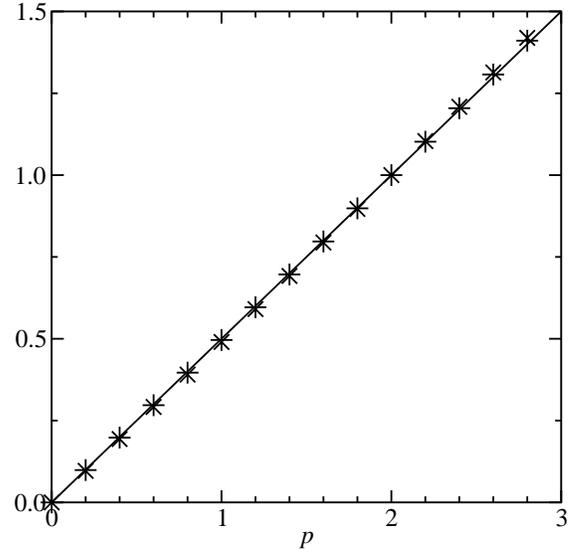}
\caption{The exponents
$\hat{\zeta}_p/\hat{\zeta}_2$~(plusses) and $\zeta^*_p/\zeta^*_2$~(crosses)
as a function of $p$ for pathological heartbeat data from
congestive heart failure patient. Both of them are close to 
the solid line of $p/2$.}
\label{fig4}
\end{figure}

\begin{figure}[thb]
\centering
\includegraphics[width=.45\textwidth]{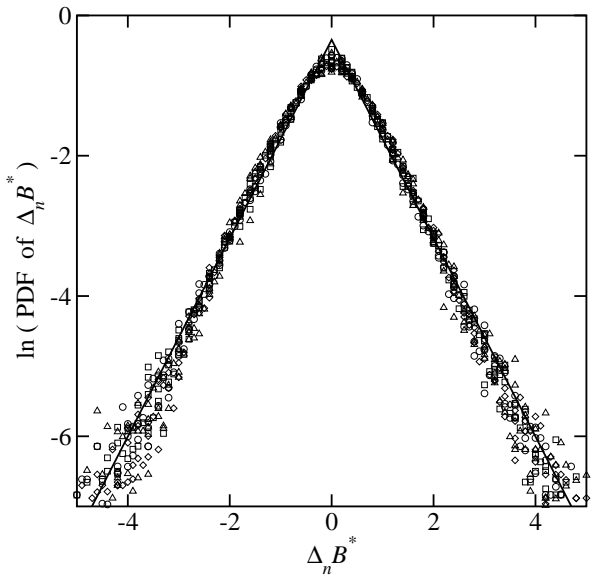}
\caption{Standardized PDFs for $\Delta_n B^*$ for healthy
heartbeat data with 
$n=4$~(circles), $n=16$~(squares), $n=64$~(diamonds),
and $n=256$~(triangles). Data from four different healthy subjects are shown and seen to coincide
with one another. These $n$-independent PDFs are seen to be well approximated by
the standard exponential distribution~(solid line).}
\label{fig5}
\end{figure}

\begin{figure}[thb]
\centering
\includegraphics[width=.45\textwidth]{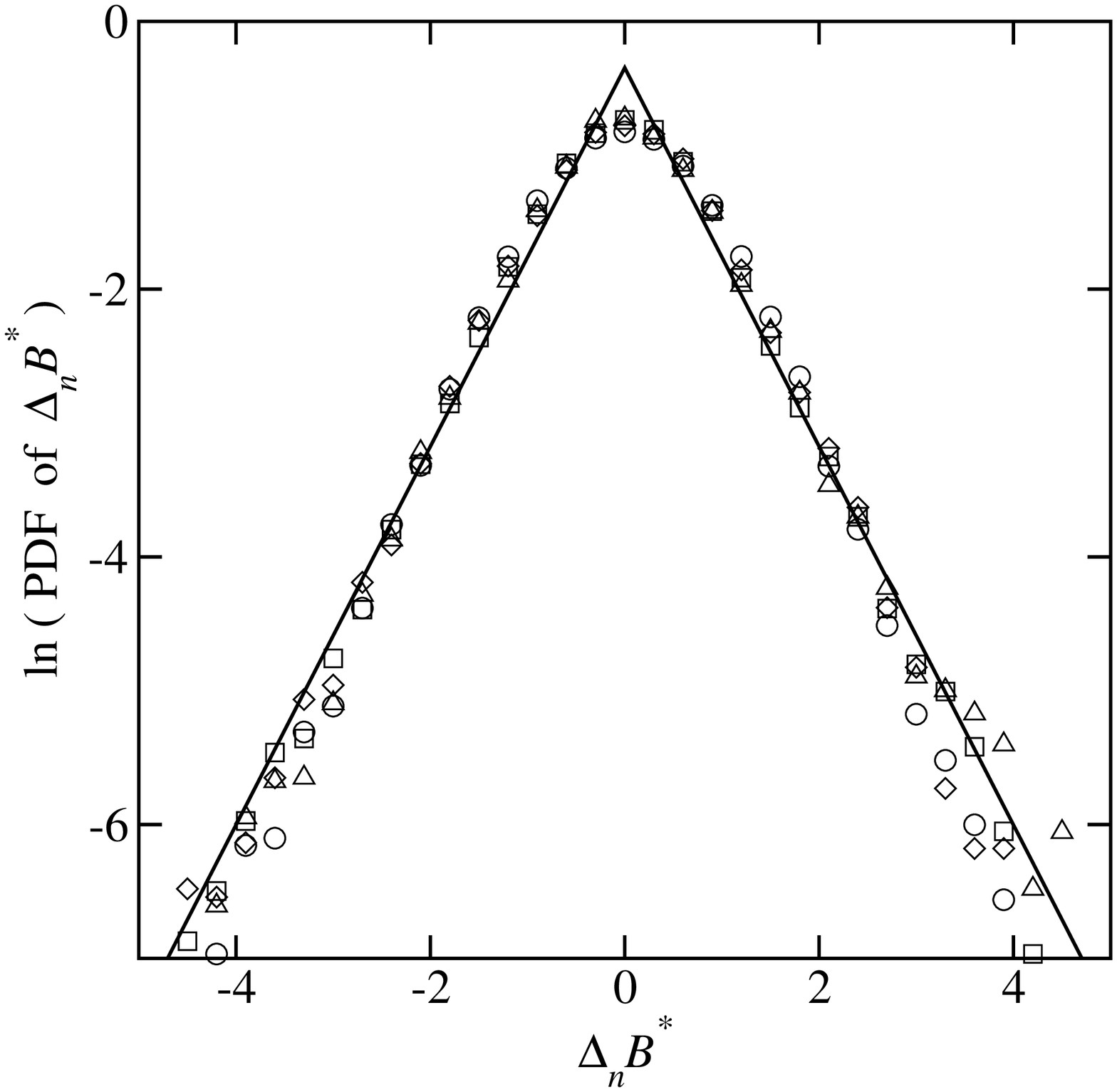}
\caption{Standardized PDFs for $\Delta_n B^*$ for
heartbeat data from a congestive heart failure patient with 
$n=4$~(circles), $n=16$~(squares), $n=64$~(diamonds), and $n=256$~(triangles).
Again, the scale invariant PDFs are well approximated by the
standard exponential distribution~(solid line).}
\label{fig6}
\end{figure}

It was suggested that this scale invariance of the standardized PDFs of $\Delta_n B^*$, 
the sum of detrended heartbeat intervals taken over $n$ beats, is an indication 
of healthy human heartbeat dynamics being in a critical state~\cite{KiyonoPRL2004}.
To check this suggestion, it would be useful to perform the same analysis to human HRV in
pathological state. Thus, we perform the same analysis using 45 sets of daytime data 
from congestive heart failure patients, also downloaded from the 
same public domain \cite{physionet}. The results for $\hat{\zeta}_p$ and
$\zeta^*_p$ in this case are shown in Fig.~\ref{fig4}.
Note that $\zeta^*_p$ is now approximately proportional to $p$,
showing that the multifractality is lost in pathological HRV, consistent with
earlier report~\cite{IvanovNature1999}.
On the other hand, $\hat{\zeta}_p$ is again proportional to $p$,
demonstrating that the scale invariance of the standardized PDFs of the sum of 
detrended heartbeat intervals is not restricted to healthy HRV but also exhibited by
pathological HRV in congestive heart failure patients. 
Moreover, we find that the scale-invariant standardized PDFs are approximately
exponential for both the healthy and pathological heartbeat 
data as shown in Figs.~\ref{fig5} and \ref{fig6}.

Since this scale invariance is found generally in heart rate fluctuations in both
healthy and pathologial state, it could not be an indication that 
healthy heartbeat dynamics are in a critical state.
Common feature for both healthy and diseased human HRV was also reported
before~\cite{BernaolaPRL2001}; it would be interesting to explore whether this
earlier feature and the present one are related.

\section{Detrended analysis for turbulent temperature measurements}

It is natural to ask whether this general scale invariance found in human heartbeat dynamics 
is trivial, i.e., whether it exists for any fluctuating data.
In this section, we shall see that such scale invariance is absent in temperature
data in turbulent flows so the answer to the above question is no. 

Specifically, we apply the detrended analysis 
to temperature measurements taken in turbulent thermal convective 
flows~\cite{Chicago}. In place of $b(i)$,
we now have $\theta(t_i)$, the temperature measurement taken at time $t_i$.
In the experiment, the measurements were sampled at a constant frequency
of 320~Hz such that $t_i = i \delta t$ with $\delta t = 1/320$~s.
The standardized PDFs of the temperature increments $\Delta_n \theta(t_i) =
\theta(t_{i+n})-\theta(t_{i})$ have been studied and found to 
change with $n$~\cite{ChingPRA} thus the temperature data in turbulent 
thermal convection are multifractal.
Also, the temperature structure functions, $R_p(n) \equiv \langle
|\Delta_n \theta(t_i)|^p \rangle$ have been studied and found to have good relative
scaling~\cite{ChingPRErapid}:
\begin{equation}
R_p(n) \sim [R_2(n)]^{\xi_p/\xi_2}
\label{xi}
\end{equation}

We calculate $\Theta(t_m) = \sum_{j=1}^m \theta(t_j)$ and 
repeat the procedure of the detrended analysis, as discussed in Sec.~III with 
$B(m)$ replaced by $\Theta(t_m)$, to obtain
$\Delta_n \Theta^*(t_i)$ and $\Delta_n \theta^*(t_i)$. We then
calculate the corresponding statistical moments $\hat{R}_p(n) 
\equiv \langle |\Delta_n \Theta^*(t_i)|^p \rangle$ and
$R^*_p(n) \equiv \langle |\Delta_n \theta^*(t_i)|^p \rangle$ 
and their respective relative exponents $\hat{\xi}_p/\hat{\xi}_2$
and $\xi^*_p/\xi^*_2$, defined by:
\begin{eqnarray}
\hat{R}_p(n) &\sim& [\hat{R}_2(n)]^{\hat{\xi}_p/\hat{\xi}_2} \\
R^*_p(n) &\sim& [R^*_2(n)]^{\xi^*_p/\xi^*_2}
\end{eqnarray}

\begin{figure}[thb]
\centering
\includegraphics[width=.45\textwidth]{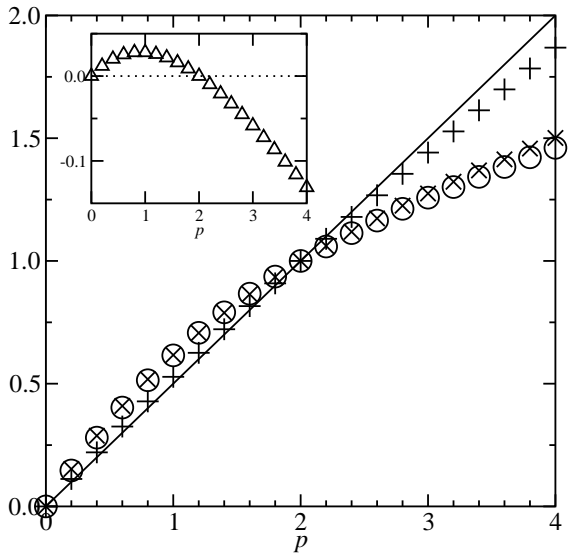}
\caption{The three relative exponents
$\hat{\xi}_p/\hat{\xi}_2$~(plusses), $\xi^*_p/\xi^*_2$~(crosses),
and $\xi_p/\xi_2$~(circles) for temperature measurements in 
turbulent convective flows. All the three relative exponents deviate from
$p/2$~(the solid line). The deviation $\hat{\xi}_p/\hat{\xi}_2 -p/2$ 
is plotted versus $p$ in the inset to show 
clearly that $\hat{\xi}_p/\hat{\xi}_2$ is not proportional to $p$.}
\label{fig7}
\end{figure}

Our results are shown in Fig.~\ref{fig7}.
Again we find that $\xi^*_p/\xi^*_2$ deviates from $p/2$, as
expected from the multifractal character of the turbulent temperature measurements. 
However, interestingly $\hat{\xi}_p/\hat{\xi}_2$ deviates from $p/2$ too, showing
that the standardized PDFs of $\Delta_n \Theta^*$ are scale dependent and changing 
with $n$. To show this deviation more clearly, we plot $\hat{\xi}_p/\hat{\xi}_2 -p/2$
versus $p$ in the inset of Fig.~\ref{fig7}.

\begin{figure}[thb]
\centering
\includegraphics[width=.45\textwidth]{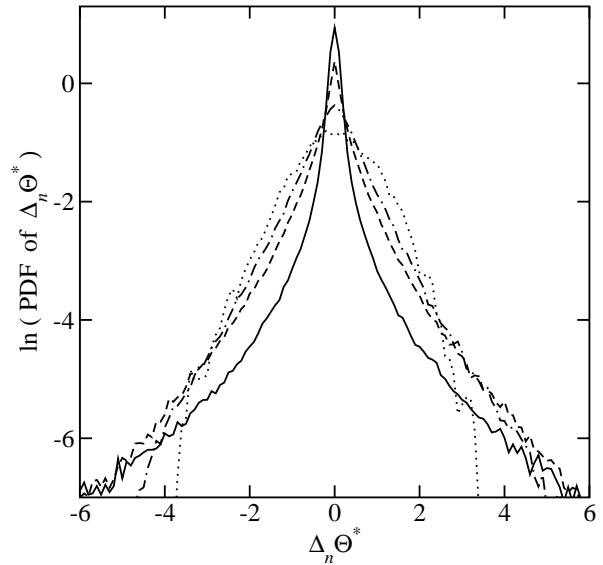}
\caption{The standardized PDFs of $\Delta_n \Theta^*$ for
temperature measurements taken in turbulent thermal convective flows
with $n=4$~(solid), $n=32$~(dashed), $n=256$~(dot-dashed) and $n=4096$~(dotted). 
The dependence of the standardized PDFs on $n$ is clearly seen.}
\label{fig8}
\end{figure}

Indeed, the standardized PDFs of $\Delta_n \Theta^*$ changes from
stretched-exponential to exponential to Gaussian as $n$ increases from 4 to 4096,
as shown explicitly in Fig.~\ref{fig8}. This change of the standardized PDFs of
$\Delta_n \Theta^*$, the sum of detrended temperature measurements taken over $n$
sampling intervals, with $n$ is similar to the change of the standardized PDFs of 
the temperature increments $\Delta_n \theta$ with $n$ as reported in
Ref.~\cite{ChingPRA}.

As can be seen in Fig.~\ref{fig7}, 
$\xi^*_p/\xi^*_2$ are close to $\xi_p/\xi_2$, indicating again
that the detrended procedure does not affect the scaling exponents of the 
temperature increments. We shall understand this in the next section.

\section{The essential effect of the detrended analysis}

In this section, we shall explore and understand what the detrended procedure
does to the data. As discussed in Sec.~III, the ``trend" in the data
is estimated by a polynomial fit in each segment of the dataset of $B(m)$, 
and we have used a polynomial of degree 3. We check that our results do not change much when
a polynomial of a lower degree is used instead. In particular, we obtain similar results
by using a linear fit of the different segments of $B(m)$. In the following,
we shall get explicit results for ``detrended" $B^*$ when the ``trend" is estimated
by a linear fit.

Let us focus on the $l$th segment of  
$B(m)$ with $m_1 \le m \le m_2$, where $m_1=(l-1)(2n)+1$ and $m_2=l(2n)$ for some $l$. 
$l$ runs from $1,2,3, \ldots$ for all the segments.
Denote the best linear fit to this segment by $a_l \ m + c_l$ where the
fitting constants $a_l$ and $c_l$ depend on $l$. The fitting
constant $a_l$ can be reasonably well approximated by the slope in this segment:
\begin{equation}
a_l \approx \frac{B(m_2)-B(m_1)}{2n-1} 
\end{equation}
Using Eq.~(\ref{Bm}), we have
\begin{equation}
a_l \approx \frac{\sum_{j=m_1+1}^{m_2} b(j)}{2n-1} \approx 
\frac{\sum_{j=m_1}^{m2} b(j)}{2n} \equiv \bar{b}_l
\end{equation} 
where $\bar{b}_l$
is the local average of $b(j)$ in the $l$th segment.
Recall from Sec.~III that $B^*$ is $B$ subtracting the best linear fit and
use Eq.~(\ref{Bm}), we have
\begin{equation}
B^*(m) \approx 
\sum_{j=1}^{m} [b(j)-{\bar b}_l] - c_l
\end{equation}
and 
\begin{equation}
B^*(m+n) \approx 
\begin{cases}
\sum_{j=1}^{m+n} [b(j)-{\bar b}_l] - c_l & 
m+n \le m_2 \\
\sum_{j=1}^{m+n} [b(j)-b_{l+1}] - c_{l+1} & 
m+n > m_2
\end{cases}
\end{equation}
Thus using Eq.~(\ref{DeltaB*}), we get
\begin{equation}
\Delta_n B^*(m) \approx 
\sum_{j=m+1}^{m+n} [b(j)-\bar{b}_l] 
\label{B*}
\end{equation}
for $m+n \le m_2$ and
\begin{equation}
\Delta_n B^*(m) \approx 
\sum_{j=m+1}^{m_2} [b(j)-\bar{b}_l] \ + \sum_{j=m_2+1}^{m+n} 
[b(j) - \bar{b}_{l+1}] 
\label{B*2}
\end{equation}
for $m+n > m_2$. 
To obtain Eq.~(\ref{B*2}), we make use of the approximation
that the two linear fits of the $l$th and $(l+1)$th segments intersect at $m=m_2$:
\begin{eqnarray}
\nonumber
b_l \ m_2 + c_l &\approx& b_{l+1} \ m_2 + c_{l+1}  \\
\Rightarrow c_{l+1}-c_l &\approx& ({\bar b}_{l}-{\bar b}_{l+1})m_2
\end{eqnarray}
Comparing Eqs.~(\ref{B*}) and (\ref{B*2}) with (\ref{deltaB*}), we see immediately
that the detrended heartbeat interval $b^*$ is given approximately by
\begin{equation}
b^*(j) \approx b(j) - \bar{b}_l
\label{detrend}
\end{equation}
Hence what the detrended analysis essentially does is to subtract the local average 
from the data. 

To verify this directly, we redo the analysis for the heartbeat data with the 
local mean subtracted and compare the results obtained with those from 
the detrended analysis. We define
\begin{eqnarray}
{\tilde b}(j) &=& b(j) - \bar{b}_l \\
{\tilde B}(m) &=& \sum_{i=1}^{m} {\tilde b}(i)
\end{eqnarray}
and study the scaling behavior of the statistical moments of
\begin{eqnarray}
\Delta_n {\tilde b}(j) &=& {\tilde b}(j+n)-{\tilde b}(j) \\
\Delta_n {\tilde B}(j) &=& {\tilde B}(j+n)-{\tilde B}(j) 
= \sum_{i=j+1}^{j+n} {\tilde b}(i)
\end{eqnarray}
with $n$.
The corresponding exponents are denoted by $\alpha_p$ and $\hat{\alpha}_p$, which
are defined by
\begin{eqnarray}
\langle |\Delta_n {\tilde b}(i)|^p \rangle &\sim& n^{\alpha_p} \\
\langle |\Delta_n {\tilde B}(i)|^p \rangle &\sim& n^{\hat{\alpha}_p}
\end{eqnarray}

\begin{figure}[thb]
\centering
\includegraphics[width=.45\textwidth]{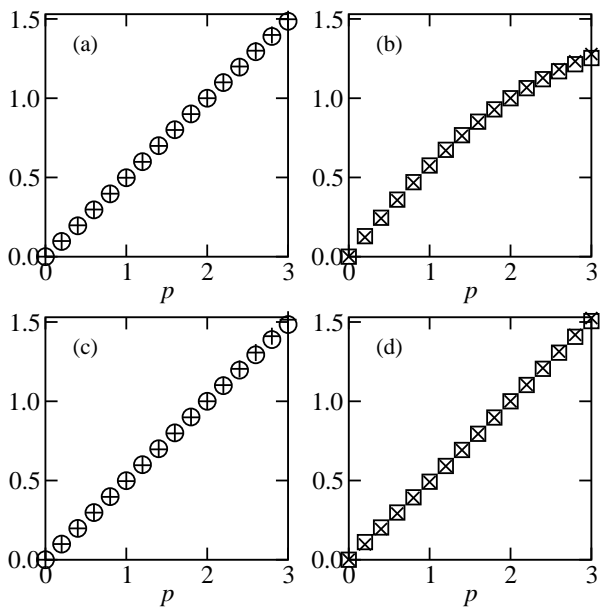}
\caption{Comparison of the exponents $\hat{\zeta}_p/\hat{\zeta}_2$~(plusses) and
$\zeta^*_p/\zeta^*_2$~(crosses) with $\hat{\alpha}_p/\hat{\alpha}_2$~(circles)
and $\alpha_p/\alpha_2$~(squares)
obtained respectively from the detrended analysis and from the
analysis eliminating the local mean. The comparison for healthy heartbeat data is shown in
(a) and (b) while that for pathological heartbeat data from congestive heart failure patients 
is shown in (c) and (d). 
Good agreement between $\hat{\zeta}_p$ and $\hat{\alpha}_p$ and between
${\zeta}_p^*$ and $\alpha_p$ is seen.}
\label{fig9}
\end{figure}

We compare $\alpha_p$ and $\hat{\alpha}_p$ with
$\zeta^*_p$ and $\hat{\zeta}_p$ respectively.
As seen from Fig.~\ref{fig9}, the results are in good agreement confirming that 
the essential effect of the detrended analysis is to eliminate 
the local average from the data. 

As a result, $b^*(j+n)-b^*(j) \approx {\tilde b}(j+n)-{\tilde b}(j)$ will be close to
$b(j+n)-b(j)$, that is, the heart rate increments are not affected much by 
the detrended analysis. This explains why $\zeta^*_p$ are close to $\zeta_p$~(see
Fig.~\ref{fig3}) and similarly why $\xi^*_p$ are close to $\xi_p$~(see
Fig.~\ref{fig7}). On the other hand, $B^*(j+n)-B^*(j) \approx {\tilde B}(j+n)-{\tilde
B}(j)$ can be different from $B(j+n)-B(j)$, and thus the 
sum of detrended heartbeat intervals could have different statistical 
features from those of the sum of untreated heartbeat intervals.

\section{Summary and Conclusions}

Understanding the nature of the complicated human HRV and thus human heartbeat dynamics
has been the subject of many studies. An interesting and intriguing finding reported in earlier 
studies~\cite{IvanovNature1999} is that in the healthy state, human 
heart rate fluctuations display multifractality, and that this multifractal 
character is lost for heart rate fluctuations in pathological state such as congested heart failure.
Based on an analogy with measurements in turbulent fluid flows, which are known to
have multifractal character as well, such multifractality in healthy HRV can be manifested
as a scale dependence of the standardized PDFs of the increment of heartbeat intervals at
different scales or between different number of beats.
A detrending analysis aiming to eliminate the non-stationarity of heartbeat data has been performed,
and "scale invariance of the PDFs of detrended healthy human heart rate increments"  
reported~\cite{KiyonoPRL2004}. We have resolved this paradox by clarifying that the scale invariance
found in the detrended analysis is actually exhibited by the PDFs of the
sum of detrended heartbeat intervals taken over different number
of beats, and demonstrating explicitly that the PDFs of detrended healthy heart rate
increments are scale dependent. We have understood the essential effect of this detrended analysis
is to eliminate the local average from the heartbeat data. We have further found that this scale
invariance of the PDFs of the sum of heartbeat intervals, with the local mean subtracted, is 
displayed also by heart rate fluctuations of congestive heart failure patients. In both the healthy
and pathological states, such scale-invariant PDFs are close to an exponential distribution. 
On the other hand, this scale invariance is absent in the multifractal 
temperature measurements in turbulent thermal
convective flows. Hence we have found an interesting scale invariance of exponential PDFs 
in human heartbeat dynamics, which is exhibited generally by heart rate fluctuations 
in both healthy and pathological states. Since this scale invariance is a general feature, 
it cannot be an
indication of the healthy state being critical, in contrast to what was claimed in earlier 
studies~\cite{KiyonoPRL2004}.

\acknowledgments

ESCC thanks D.C. Lin for discussions in the early part of this work.
This work is supported in part by the Hong Kong Research Grants Council
(Grant No. CUHK 400304).

\end{document}